\documentclass[aps,pra,twocolumn,amsmath,amssymb,footinbib,showpacs,longbibliography,superscriptaddress]{revtex4-2}

\usepackage[english]{babel}
\usepackage{latexsym}
\usepackage{graphics}
\usepackage{subfigure}
\usepackage{epsfig}
\usepackage{xcolor}
\usepackage{hyperref}
\usepackage{braket} 
\usepackage[T1]{fontenc}
\usepackage[latin9]{inputenc}
\usepackage{amstext}
\usepackage{amssymb}
\usepackage{graphicx}
\usepackage{esint}
\usepackage{braket}
\usepackage{babel}
\usepackage{amsmath}
\usepackage{siunitx}
\usepackage{autobreak}
\usepackage{comment}
\usepackage{booktabs}
\usepackage{tabularx}
\usepackage{placeins}

\hypersetup{
	colorlinks=true,
	citecolor=blue,
	linkcolor=red,
	urlcolor=black
}


\newcommand{\e}{\textrm{e}}
\newcommand{\ie}{i.e.}

\newcommand{\D}{\mathrm{d}}
\newcommand{\pa}{\partial}

\newcommand{\inftyint}{\int_{-\infty}^{\infty}}

\newcommand{\kB}{k_{\mathrm{B}}}
\newcommand{\kF}{k_\textrm{\tiny F}}
\newcommand{\vF}{v_\textrm{\tiny F}}

\newcommand{\Ec}{E_c}
\newcommand{\TF}{T_\textrm{\tiny F}}
\newcommand{\U}{U_0}
\newcommand{\Rc}{R_c}

\begin{document}

\title{
Phase diagram of one-dimensional bosons with Rydberg-dressed soft-core interactions
}

\author{Shengjie Yu}
\affiliation{CPHT, CNRS, \'Ecole Polytechnique, Institut Polytechnique de Paris, Palaiseau, France}

\author{Laurent Sanchez-Palencia}
\affiliation{CPHT, CNRS, \'Ecole Polytechnique, Institut Polytechnique de Paris, Palaiseau, France}

\date{\today}

\begin{abstract}
Rydberg and Rydberg-dressed atomic gases have recently emerged as a promising quantum simulator for a variety of models in condensed matter physics.
Here we investigate one-dimensional bosons with soft-core Rydberg-dressed interactions using exact path-integral quantum Monte Carlo simulations. The finite-range and the negative Fourier component of the interaction potential generate a roton mode at finite momentum, while particle-hole backscattering processes enhance the susceptibility of one-dimensional systems at twice the Fermi momentum. The competition between the corresponding length scales yields a rich phase diagram, featuring a conventional Tomonaga-Luttinger liquid (TLL) regime, a beyond-TLL regime, and commensurate cluster phases. In the TLL regime, the system transitions from Lieb-Liniger-like bnehavior with Luttinger parameter $K>1$ to hard-rod-like behavior with $K<1$, with a quasi-supersolid phase emerging for $K < 1/2$. For strong interactions and high densities, deviations from TLL theory appear as precursors for the onset of cluster phases, where particles aggregate into stable clusters of several particles. The properties of each phase is discussed in detail.
\end{abstract}

\maketitle

\section{Introduction} \label{sec:introduction}
Neutral atoms are a surging platform of quantum computation and simulation, owing to their strong interatomic interactions and high control capability~\cite{lewenstein2007,bloch2008,bloch2012,gross2017,esslinger2010,tarruell2018,*lsp2018,weimer2010,saffman2010quantum,browaeys2020many,altman2021quantum}.
A widely exploited property is the Rydberg blockade, which prevents the simultaneous excitation of neighboring atoms~\cite{jaksch2000fast,lukin2001dipole,urban2009,gaetan2009} and has been extensively used for implementing quantum gates~\cite{jaksch2000fast,lukin2001dipole,isenhower2010demonstration}.
Alternatively, by weakly coupling atoms from the ground state to a Rydberg state using a laser, one can engineer a soft-core interaction potential~\cite{johnson2010interactions,honer2010collective}. The strength and characteristic range of this potential can be tuned via the laser intensity and detuning, while at large distances, the long-range interaction inherits the properties of the bare inter-atomic van der Waals interaction.
Interestingly, in momentum space, the interaction exhibits a negative component, which gives rise to a roton mode in the excitation spectrum when the interaction strength is sufficiently large~\cite{santos2003,lahaye2009}.
The latter favors density modulations with a wavelength set by the roton momentum~\cite{baranov2012condensed}. Closing the roton gap signals an instability of the homogeneous phase toward finite-momentum density order, which may manifest as density waves, supersolid or crystalline order, cluster crystals, or droplet arrays, depending on dimensionality and microscopic models.~\cite{leggett1970can,santos2003,lahaye2009,mottl2012roton,petrov2015,ferrier2016observation,schmitt2016self,tanzi2019supersolid,bottcher2019transient,hertkorn2021density}.

In one-dimensional (1D) homogeneous quantum systems, genuine crystallization does not occur due to the inhibition of spontaneous breaking of continuous symmetry~\cite{mermin1966}.
Further peculiarities of physics in one dimension include strong phase fluctuations, absence of genuine long-range order, and breakdown of the quasiparticle picture, 
making 1D systems significantly different from their counterparts in higher dimensions~\cite{giamarchi2003}.
So far, a wide variety of effective 1D systems has been identified, including organic conductors, carbon nanotubes, confined ultracold atoms, as well as edge modes of quantum Hall materials~\cite{giamarchi2003,cazalilla2011}.
In such systems, the conventional Fermi liquid theory is not applicable.
Instead, Tomonaga-Luttinger liquid (TLL) theory, incorporating collective modes, successfully describes most gapless short-range 1D systems~\cite{cazalilla2004bosonizing}.
The TLL theory predicts quasi-long-range order, characterized by an algebraic decay of correlation functions at large distances, with the decay exponent being governed by the Luttinger parameter $K$.
Moreover, strong repulsive interactions enhance antibunching and a tendency towards formation of charge-density waves with a lattice constant given by the inverse density, induced by particle-hole backscattering processes~\cite{giamarchi2003,cazalilla2011}.
Besides, structured interactions may induce a gapless roton mode, which challenge TLL theory.
Moreover, the cooperation of charge-density-wave enhancement and roton gap closing with a characteristic momentum commensurate with
the average particle distance, has been shown to induce the emergence of clustering~\cite{rossotti2017quantum,mambretti2020emergence,prestipino2019clusterization,teruzzi2021evolution}.
Particles then form molecule-like clusters, and the clusters subsequently arrange into an effective TLL.
By varying the interaction strength, an emergent Ising-type critical point is identified at the transition point of the two-particle cluster phase.
However, cluster formation in regimes away from commensurability remains largely unexplored.
Certain studies of classical soft-core particles report signatures of a sharp crossover, reminiscent of a first-order phase transition~\cite{prestipino2014cluster,prestipino2015probing} while meanfield studies on quantum systems imply a continuous quantum phase transition~\cite{prestipino2019clusterization}.

\begin{figure}[t]
\centering
\includegraphics[width=\linewidth]{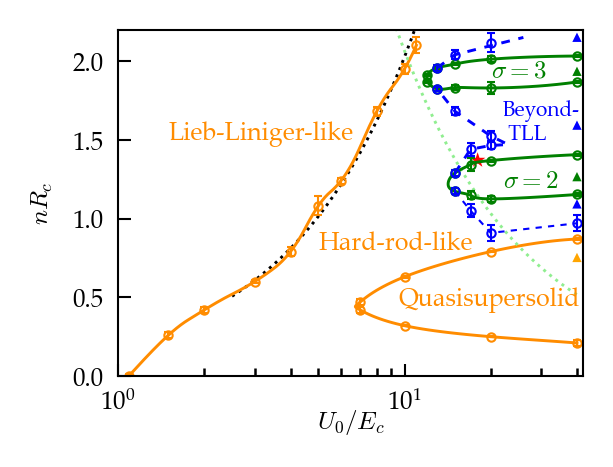}
\caption{Quantum phase diagram for bosons with Rydberg-dressed soft-core interactions versus interaction strength and density (semi-logarithmic scale).
The solid orange lines indicate the thresholds between
Lieb-Liniger-like ($K>1$), hard-rod-like ($1/2<K<1$) and quasi-supersolid  ($K<1/2$) behaviors.
The dashed blue lobes indicate beyond-TLL regions and the solid green lines the cluster lobes, where $\sigma$ denotes the particle number in each cluster.
The dotted lines indicate meanfield Bogoliubov predictions for the $K=1$ line (dotted black) and the line of roton-gap closing (dotted green).
The red star shows the phase transition point at commensurate roton and density length scales, identified in Ref.~\cite{rossotti2017quantum}.
}
\label{fig:phase_diagram}
\end{figure}

In this work, we study 1D bosons with the soft-core interaction potential $U(x) = \U / [1 + (x/\Rc)^6]$ using exact path-integral quantum Monte Carlo (QMC) calculations in continuous space.
Such an interaction potential can be engineered via Rydberg dressing, the parameters of which (interaction strength $\U$ and characteristic range $\Rc$) are tunable through the laser detuning and intensity~\cite{henkel2010three,honer2010collective}.
For sufficiently large density and interaction strength, this potential induces a roton mode in the excitation spectrum~\cite{rossotti2017quantum,prestipino2019clusterization}.
We show that the competition between
particle-hole backscattering processes and roton-driven ordering leads to rich and fascinating phenomena.
Our main results are summarized in the phase diagram presented in Fig.~\ref{fig:phase_diagram} versus the interaction strength $\U$ and the density $n$.
Roughly speaking, when the Fermi energy of the 1D gas exceeds the low-momentum interaction energy per particle, the details of the soft-core interaction become irrelevant, and the system effectively behaves as a Lieb-Liniger gas.
This regime is characterized by a Luttinger parameter $K > 1$, transitioning from a weakly interacting regime for $K \gg 1$ to a strongly interacting regime as $K$ approaches the Tonks-Girardeau limit $K \rightarrow 1$.
In the opposite case, where the potential core acts as a strong barrier, the gas behaves as an effective hard-rod (HR) gas, with $K < 1$. For sufficiently strong interactions and moderate density, the system further enters a regime where $K<1/2$ and the structure factor exhibits diverging peaks, signaling the onset of quasi-long-range diagonal order and solidification-like behaviour. Nevertheless, the Rydberg-dressed atomic system remains a 1D superfluid characterized by quasi-long-range off-diagonal order, thus forming a kind of quasi-supersolid phase. Throughout this region of the phase diagram, our QMC results are indeed consistent with TLL behaviour.
In contrast, for sufficiently strong interactions and high density, we observe a breakdown of TLL theory, corresponding to the beyond-TLL region enclosed by the dashed blue line in the phase diagram. This appears as a precursor to cluster phases, where particles aggregate into clusters containing a fixed number of particles, see the green lobes. Our results are consistent with, yet significantly extend, previous findings. The red star marks the critical point for clustering at commensurate roton and density length scales, as identified in Ref.~\cite{rossotti2017quantum}. Here, we find that cluster phases are stabilized away from exact commensurability.
Meanfield theory predicts that the closing of the roton gap, depicted as a dotted green line in the diagram, leads to supersolid density modulations in the ground state.
While this mechanism is consistent with the tips of the cluster lobes, it fails to predict the unclustered quasi-supersolid phase found from QMC calculations.

In the following, we first introduce the model and methods (Sec.~\ref{sec:model_methods}), then discuss the physical properties of the Rydberg gas in the various regimes (Secs.~\ref{sec:normal_TLL} and \ref{sec:beyondTLL}),
and finally present our conclusions  (Sec.~\ref{sec:conclusions}).

\section{Model and Methods} \label{sec:model_methods}
\begin{figure}[t]
\centering
\includegraphics[width=\linewidth]{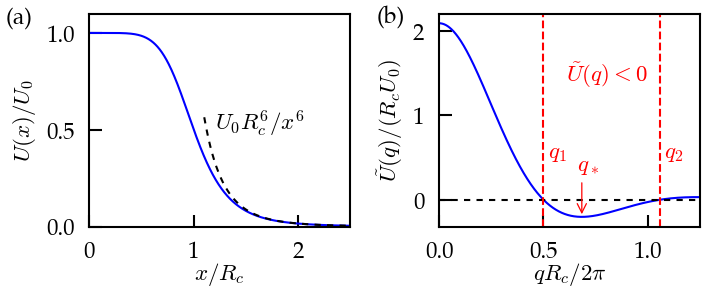}
\caption{Interaction potential in (a)~real space and (b)~momentum space, indicated by solid blue lines. 
The dashed lines represent the van der Waals tail in~(a) and the negative range of $\tilde{U}(q)$ in~(b), with the arrow indicating the minimum of $\tilde{U}(q)$ at $q=q_*$.}
\label{fig:interaction}
\end{figure}

The Hamiltonian for spinless bosons in homogeneous 1D space under periodic boundary conditions reads as
\begin{equation} \label{eq:hamiltonian}
\hat{H} = \sum_{j=1}^N \frac{\hat{p}_j^2}{2M} + \sum_{j<\ell} U(\hat{x}_j - \hat{x}_\ell),
\end{equation}
where $M$ denotes the particle mass
and
the interaction potential, $U(x) = \U / [1 + (x/\Rc)^6]$, is that of Rydberg-dressed atoms~\cite{henkel2010three,honer2010collective}. 
In real space, it is characterized by a soft core plateau with energy $\U=U(x=0)$, a sharp shoulder around $x\approx \Rc$, and a long-range van der Waals tail for $x\gg \Rc$, see Fig.~\ref{fig:interaction}(a).
In momentum space, its Fourier transform,
\begin{equation} \label{eq:potentialq}
\begin{aligned}
\tilde{U}(q) = \frac{\pi}{3} \U \Rc \e^{-q\Rc/2} &\left[ \e^{-q\Rc/2} + \right. \\ \cos\left(\frac{\sqrt{3}}{2}q\Rc\right) + &\left. \sqrt{3}\sin\left(\frac{\sqrt{3}}{2}q\Rc\right) \right],
\end{aligned}
\end{equation}
has a negative region between
$q_1 \approx 2\pi\times 0.500\Rc^{-1}$ and $q_2 \approx 2\pi\times 1.058\Rc^{-1}$,
with minimum at $q_* \approx 2\pi\times 0.684\Rc^{-1}$, see Fig.~\ref{fig:interaction}(b).
As discussed below, this region generates a roton minimum in the energy spectrum.

We study the Rydberg-dressed system by performing path-integral QMC simulations in continuous space within the grand-canonical ensemble~\cite{ceperley1995path}.
In brief, we sample a large number of world-line configurations, under given values of interaction strength $\U$, range $\Rc$, chemical potential $\mu$, and temperature $T$,
and compute average values of thermodynamic quantities including the particle number, the energy, and their fluctuations.
This directly yields the particle density $n(\mu)$ and the compressibility $\kappa \equiv \pa n/\pa \mu$. 
The worm algorithm also enables us to obtain correlation functions and extract information from them, such as the TLL parameter $K$.
Here we compute (i)~the single particle correlation function $g_1(x) = \frac{1}{n} \langle \hat{\psi}^\dagger(x) \hat{\psi}(0) \rangle$ with $\hat{\psi}(x)$ the Bose field operator at position $x$,
(ii)~the pair correlation function $g_2(x) = \frac{1}{n^2} \langle \hat{\psi}^\dagger(0) \hat{\psi}^\dagger(x) \hat{\psi}(x) \hat{\psi}(0) \rangle$,
and (iii)~the static structure factor $S(q) = [\langle \rho_q \rho_{-q} \rangle - |\langle \rho_q\rangle|^2] / N$, with $\rho_q = \int dx\,n(x) e^{i q x}$,
as obtained via the Fourier relation $S(q) = 1 + n \inftyint \D x\,[g_2(x)-1] e^{iqx}$~\cite{pitaevskii2016bose}.
In order to jump among different particle number sectors, the QMC algorithm incorporates open worldlines (worms), whose two ends exactly map the creation and annihilation operators, hence allowing computing $g_1(x)$.
In contrast, for a sample without worms, diagonal terms are measured, including the particle number $N$, and correlation functions $g_2$ and $S(q)$.
After gathering more than $10^5\sim 10^6$ samples, the statistical fluctuations are usually sufficiently small such that we can neglect the errorbars in following computations.
In most computations, the system size is $L=500\Rc$, such that finite size effects are negligible.

\begin{figure}[t]
\centering
\includegraphics[width=\linewidth]{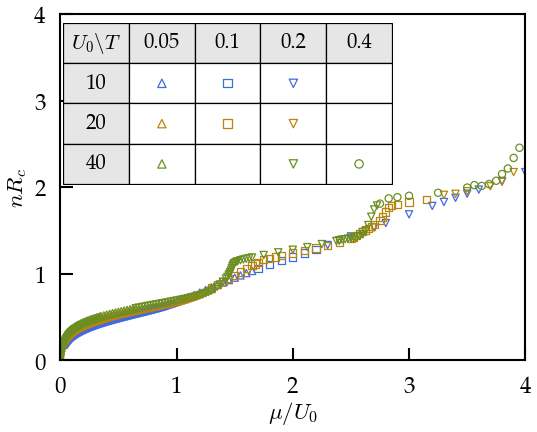}
\caption{Equation of state for 1D bosons with Rydberg-dressed soft-core interactions at various interaction strengths $\U$ and negligible temperatures, as computed using QMC simulations.
The various interaction strengths are encoded with different colors and the various temperatures with different marker shapes, with values indicated in the table.
Errorbars are smaller than the markers.
}
\label{fig:eqnofstates}
\end{figure}

The equation of state $n(\mu)$ is displayed in Fig.~\ref{fig:eqnofstates}  for various values of $\U$, with the caption listing all temperatures.
In QMC, the strength $\U$ is fixed and the temperature is set to be negligible
(\ie~such that $T\ll \TF$ where $\TF = \hbar^2 \kF^2 / 2M\kB$ is the Fermi temperature and $\kF=\pi n$ is the Fermi wave number in one dimension),
except for a few points around $\mu \approx 0$.
The results clearly show different behaviours depending on the interaction strength.
For $\U=10\Ec$, with energy unit $\Ec=\hbar^2/M\Rc^2$, we find that the density grows smoothly with the chemical potential. The behavior of $n(\mu)$ interpolates between $n \propto \sqrt{\mu}$ for $\mu \rightarrow 0$, characteristic of the vacuum-to-TLL quantum phase transition~\cite{sachdev2001}, and $n \propto \mu$ in the limit $\mu \rightarrow \infty$, characteristic of the meanfield regime. This behavior is reminiscent of that observed in standard 1D Bose gases, such as the LL and HR models, see for instance Ref.~\cite{yu2026thermodynamics}.
In contrast, for larger values of $\U$, we observe jumps in the density, which become increasingly sharp as $\U$ increases.
We observe that the jumps occur around specific densities and chemical potentials,
namely $n\Rc \simeq 0.95$, $1.65$, and $2.5$
and
$\mu/\U \simeq 1.4$, $2.6$, and $3.8$, with roughly equal spacing.
Between the jumps, the density increases linearly with the chemical potential, corresponding to plateaus in the compressibility.
This behavior signals that the system undergoes fundamentally distinct physical processes, which are strongly dependent on both the density and the interaction strength.
By comparing Fig.~\ref{fig:eqnofstates} with the phase diagram in Fig.~\ref{fig:phase_diagram}, we find that the cluster phases correspond precisely to the plateaus observed in the compressibility $\kappa$.
However, the equations of state alone are insufficient to reveal the specific phases.
In the following, we provide a detailed characterization of the physical properties of the various regimes identified in the phase diagram of Fig.~\ref{fig:phase_diagram}.

\section{Normal Tomonaga-Luttinger liquid} \label{sec:normal_TLL}

\begin{figure*}[t]
\centering
\includegraphics[width=\linewidth]{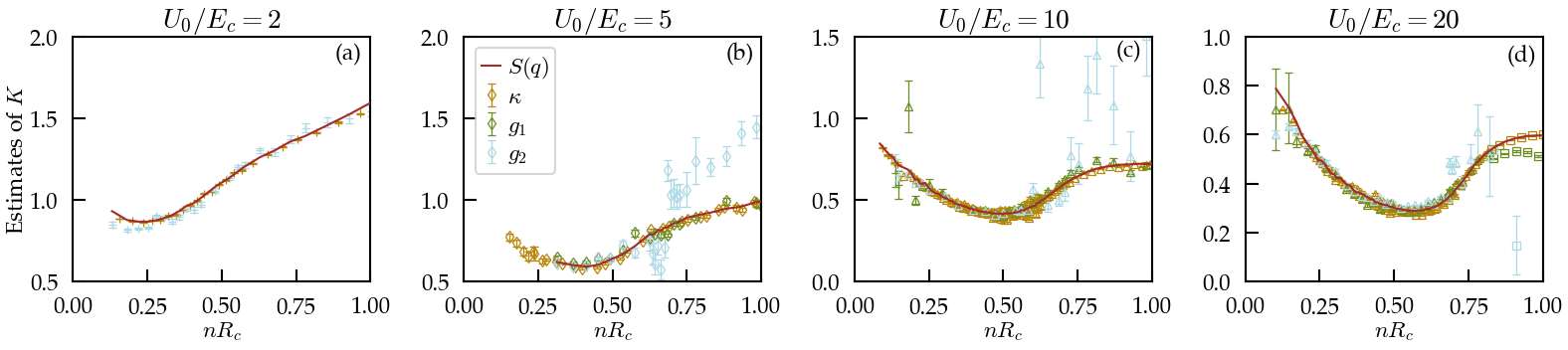}
\caption{Luttinger parameter $K$ versus density of the Rydberg-dressed 1D gas for various values of the interaction strength $\U$. The values of $K$ are estimated from the compressibility $\kappa$, the correlation functions $g_1$ and $g_2$, and the structure factor, see text.}
\label{fig:K_normalTLL}
\end{figure*}

We start with the normal TLL regime.
In zero-temperature TLL theory, one considers a low-energy linear dispersion relation of collective modes, characterized by the speed of sound $u$ and leading to algebraic decay of correlation functions in the long-distance limit.
Although most of our calculations are performed at sufficiently low temperatures for thermodynamic quantities to be effectively evaluated at zero temperature, the long-distance behavior of correlation functions is always affected by thermal fluctuations beyond the thermal distance $L_T=\hbar u/\kB T$.
At finite temperatures, the asymptotic behavior of correlation functions is then described by modified Haldane's formulas~\cite{haldane1981effective,haldane1981luttinger,giamarchi2003,cazalilla2004bosonizing},
\begin{equation} \label{eq:TLL_g1}
g_1(x) = \frac{1}{|nd(x)|^{1/2K}} \sum_{m=0}^\infty B_m\frac{\cos(2\pi m nx)}{|nd(x)|^{2m^2K}},
\end{equation}
and
\begin{equation} \label{eq:TLL_g2}
g_2(x) = 1 - \frac{2K}{|2\pi nd(x)|^{2}} + \sum_{m=1}^\infty A_m \frac{\cos(2\pi m n x)}{|nd(x)|^{2m^2K}},
\end{equation}
where $d(x)=\frac{L_T}{\pi}\sinh(\frac{\pi x}{L_T})$ is the thermal chord distance, and the coefficients $A_m$'s and $B_m$'s are nonuniversal, model-dependent quantities.
Hence, the correlation functions decay algebraically for $x \ll L_T$ and exponentially for $x \gg L_T$, with oscillations terms at frequencies $f_m=m\times n$.

Our QMC calculations provide us with four different estimates of the Luttinger parameter $K$.
First, we fit the Haldane formulas to both correlation functions $g_1(x)$ and $g_2(x)$ computed by QMC, and extract two estimates of the Luttinger parameter $K$ and the `density'' $n$ from the oscillation frequency $f_1$, as well as the coefficients $A_1$, $B_0$ and $B_1$, with higher order terms being neglected. For more details of the fits, see Ref.~\cite{yu2026thermodynamics}.
The fit range of $x$ is selected manually to ensure a good signal-to-noise ratio.
The accessible range is typically $1<nx<100$, although its precise upper limit depends on the statistical accuracy of the QMC data.
At larger distances, statistical fluctuations tend to dominate the signal.
Second, the structure factor yields another estimate of the Luttinger parameter. In the Feynman approximation, which assumes a single-mode spectrum with energy $\varepsilon(q)$, it reads as
$S(q\to 0) \approx \varepsilon_0(q)\coth[\varepsilon(q)/2\kB T]/ \varepsilon(q)$,
where $\varepsilon_0(q) = \hbar^2 q^2/2M$ is the free-particle energy spectrum~\cite{derosi2023correlation}.
It reduces to $S(q) = {\hbar q}/{2Mu} = Kq/2\kF$ for $q\to 0$ at zero temperature for TLL,
where we have used the relation for Galilean-invariant systems, $u=\vF / K$, with $\vF=\kF/M$ the Fermi velocity and $\kF=\pi n$ Fermi wave number.
Fitting this formula to the low-$q$ behavior of $S(q)$, we extract a new value for $K$ and corresponding speed of sound $u$.
Third, using the relation $\kappa=n/Mu^2$, valid at zero temperature, we get yet another estimate of $K$ from the QMC calculated compressibility via the relation $K = \pi \hbar \sqrt{n\kappa / M}$.
It works within errorbars when the temperature is negligible.
In conclusion, we have four different estimates of $K$ from fits of $g_1$, $g_2$, and $S(q)$, and from the compressibility $\kappa$.

\begin{figure*}[t]
\centering
\includegraphics[width=\textwidth]{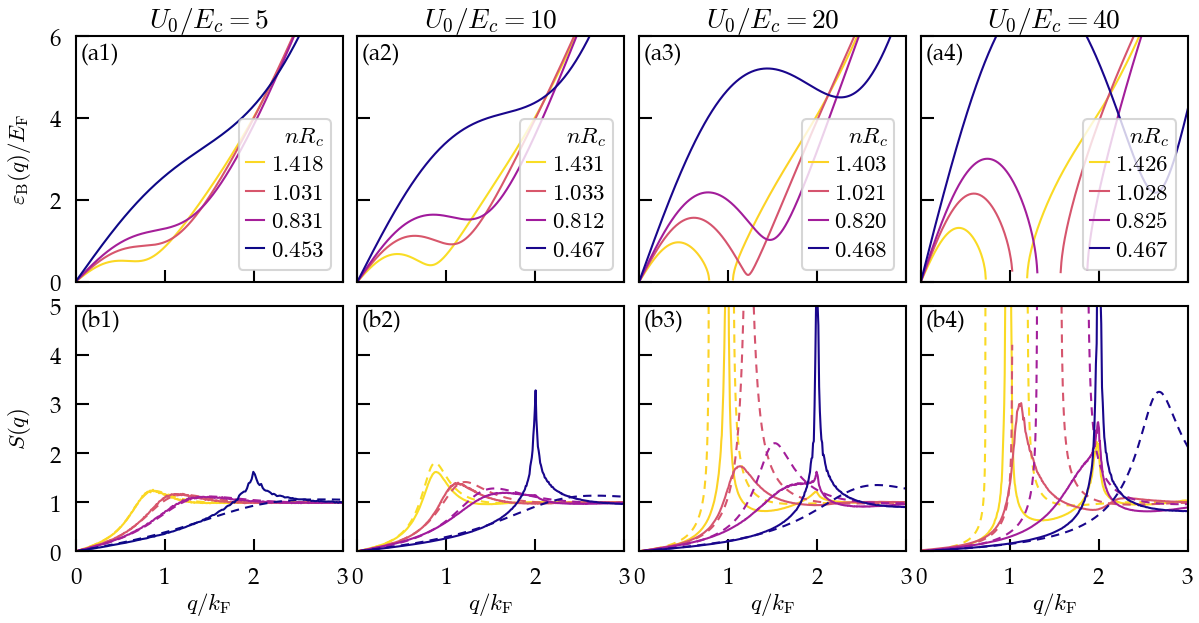}
\caption{(a)~Bogoliubov spectrum within Feynman approximation (solid lines)
and (b)~static structure factor vs $q/\kF$ from QMC (solid lines) and Feynman approximation (dashed lines),
for $\U$ ranging from $5\Ec$ to $40\Ec$ (from left to right columns).
The color lines, ranging from blue to yellow, indicate increasing densities, with the lower row sharing the same color labels as the upper row.}
\label{fig:structurefactor}
\end{figure*}

\subsection{Tomonaga-Luttinger regime}
\label{sec:TLL_regime}
A key feature of TLLs is that a unique Luttinger parameter $K$ controls the interaction regime and governs all physical properties.
Hence the coincidence of the various estimates of $K$ is a strong evidence of TLL behaviour. 
Figure~\ref{fig:K_normalTLL} shows the estimates for $K$ for various interaction strengths $\U$, ranging from $2\Ec$ to $20\Ec$,
using fits of $S(q)$, $\kappa$, $g_1(x)$, and $g_2(x)$ from QMC computations.
For low interaction strength, we find good agreement between all estimates, see Fig.~\ref{fig:K_normalTLL}(a).
For intermediate interaction strength, Figs.~\ref{fig:K_normalTLL}(b) and (c) as well as the largest part of Fig.~\ref{fig:K_normalTLL}(d), we still find good agreement, except for $K_{g_2}$ for relatively high densities.
While, in principle, it signals the breakdown of TLL theory, here this is not the case.
In practice, $g_2(x)$ decays rapidly and its values are soon dominated by QMC statistical fluctuations.
As a consequence, the available fitting range is too narrow (only up to $nx \sim 10$) to allow for a reliable fit.
On the other hand, the other three estimates, $K_\kappa$, $K_{S(q)}$, and $K_{g_1}$, are consistent with one another and reliably probe large-distance properties. 
It is therefore reasonable to conclude that the system still belongs to the TLL universality class.
In contrast, for large-enough interaction strength and density, we find a clear breakdown of TLL theory with significant deviation of the fitted values of $K$ from $g_1$ on the one hand and from $\kappa$ and $S(q)$ on the other hand, see Fig.~\ref{fig:K_normalTLL}(d) for $\U=20$ and $n\Rc \gtrsim 0.8$. This regime is discussed in more detail in Sec.~\ref{sec:beyondTLL}.
Using similar analysis for other parameters, we find that the normal TLL phase occupies most of the phase diagram in Fig.~\ref{fig:phase_diagram}, except the blue and green lobes.

Within the TLL regime, we identify three distinct subregimes based on the value of the Luttinger parameter $K$.
At sufficiently high densities and moderate interaction strengths, the Rydberg-dressed gas exhibits properties similar to the Lieb-Liniger gas, characterized by $K > 1$.
It crosses over from the weakly interacting regime ($K \gg 1$) to the strongly interacting regime ($K \simeq 1$) as the density decreases or the interaction strength increases.
For $K = 1$, the system corresponds to the Tonks-Girardeau gas, which is the infinite-interaction limit of the Lieb-Liniger gas.
For larger interaction strengths, the potential core becomes prominent and acts as a strong barrier, leading the system to behave as an effective HR gas with $K < 1$. This regime is characterized by a tendency towards formation of charge-density-wave modulations, as evidenced by slower decay of oscillations in the pair correlation function $g_2(x)$ and enhancement of peaks in the structure factor $S(q)$ at $q = 2\kF$, see blue curve in Fig.~\ref{fig:structurefactor}(b1).
For even stronger interactions, the system enters a regime with $K < 1/2$. In this regime, the structure factor $S(q)$ exhibits algebraically diverging peaks, signaling the onset of solidification-like behavior, see blue curves in Figs.~\ref{fig:structurefactor}(b3) and (b4).
Consistently, this structural order is characterized by oscillations in $g_2(x)$ that decay algebraically, corresponding to  quasi-long-range diagonal order.
Nevertheless, the gas remains a 1D superfluid, characterized by algebraic decay of the one-body correlation function, \ie~quasi-long-range off-diagonal order.
We thus refer to this regime as quasi-supersolid, where `quasi' refers to both diagonal and off-diagonal orders.
Note that for Fig.~\ref{fig:structurefactor}(b2), we have $K<1/2$ and we find a strong peak at $q=2\kF$. However, it does not strictly diverge owing to finite-temperature effects~\cite{yu2026thermodynamics}.

\subsection{Meanfield regime}
The behavior of the structure factor can be understood in the weakly-interacting regime, using meanfield approach.
In this regime, the ground state is expected to be a quasicondensate and we can apply the Bogoliubov approach~\cite{popov1972}.
The dispersion relation reads as $\varepsilon_B(q) = \sqrt{\varepsilon_0(q) [\varepsilon_0(q) + 2n\tilde{U}(q)]}$,
where $\varepsilon_0(q) = \hbar^2 q^2 / 2M$ is the free-particle dispersion and $\tilde{U}(q)$ is the interaction potential in momentum space.
For the Rydberg-dressed interaction potential given in Eq.~(\ref{eq:potentialq}), it predicts the emergence of a roton mode, \ie~a local minimum in the dispersion relation at a finite wave number. This occurs when $n\U$ is sufficiently large, allowing the minimum of $n\tilde{U}(q)$ to overcome the increase of $\varepsilon_0(q)$ with $q$. This condition is satisfied for $n\Rc \gtrsim 6.12\Ec/\U$.
The static structure factor is then computed using Feynman approximation as $S_\mathrm{FA}(q) = \varepsilon_0(q) / \varepsilon_B(q)$.
The meanfield regime is identified by comparing the static structure factor $S(q)$ as obtained from QMC with that obtained from the Bogoliubov-Feynman approximation.

Figure~\ref{fig:structurefactor}(a) shows the Bogoliubov dispersion relation (solid lines) and (b)~the structure factor $S(q)$ from QMC (solid lines), together with the Feynman approximation based on the Bogoliubov dispersion relation (dashed lines), across a large range of interaction strength $\U$ and density $n$.
We begin with low interaction, $\U=5\Ec$, corresponding to the first column.
Except for the lowest density, the higher three are clearly in the meanfield regime, for which we find good agreement between $S(q)$ from QMC and Bogoliubov-Feynman approximation.
For sufficiently high density, a roton mode appears and develops a local maximum in $S(q)$, see yellow curve, corresponding to $n\approx 1.418\Rc^{-1}$.
Because the roton is gapped in this case, the maximum in $S(q)$ is Lorenzian-like, corresponding to an exponential decay of oscillation in real space, with frequency equal to the roton wave number.
Since the decay is exponential, the long-distance algebraic TLL behavior are not broken.
We then turn to stronger interaction, $\U=10\Ec$, corresponding to the second column.
There, we find small but visible discrepancies between the QMC and Bogoliubov-Feynman predictions, indicating that we are close to the boundary between meanfield and strongly-correlated regimes.
For larger interaction strength $\U$ (the two rightmost columns), the meanfield picture clearly breaks down.
From our data, we find that the meanfield regime basically corresponds to $\U \lesssim 10\Ec$ and $n\gtrsim 0.6\Rc^{-1}$.
Although a large Luttinger parameter usually indicates weak long-wavelength phase fluctuations, it is not a sufficient criterion for the validity of the Bogoliubov meanfield prediction of the full static structure factor. In the present finite-range Rydberg-dressed system, the agreement with meanfield theory is controlled not only by $K$, but also by finite-momentum correlations associated with the roton minimum, the ratio $q/2\kF$, and the proximity to clustering (see Sec.~\ref{sec:beyondTLL}).

In addition, the Luttinger parameter $K$ can also be extracted from Bogoliubov approach.
Inserting the speed of sound $u_\mathrm{B} = \sqrt{2n\tilde{U}(0) / 2M}$ into the formula $K=\pi\hbar n / Mu_\mathrm{B}$, we find $K_\mathrm{B}=\pi\Rc \sqrt{n\Ec/\tilde{U}(0)}$.
Extrapolating to strong interactions, it yields an estimate of the Tonks-Girardeau limit, $K=1$, at $n \simeq \tilde{U}(0)/\pi^2\Ec\Rc^2$.
The latter is indicated by the black dotted line in Fig.~\ref{fig:phase_diagram}, which yields a good estimate for $3 \lesssim \U/\Ec \lesssim 10$.

\subsection{Strongly-correlated regime}
\begin{figure}[t]
\centering
\includegraphics[width=\linewidth]{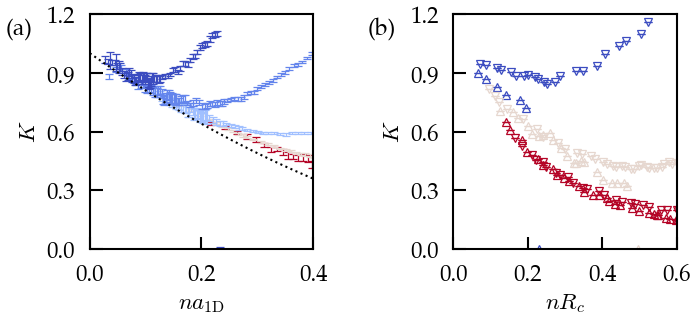}
\caption{Luttinger parameter versus density in the low-density regime.
(a)~$K$ as found from $\kappa$ for $\U/\Ec=2$, $3$, $5$, $10$ and $40$ (from blue to red) vs $na_\mathrm{1D}$, and comparison to the HR relation $K=(1-na_\mathrm{1D})^2$.
(b)~Comparison between the values of $K$ obtained for Rydberg-dressed interaction (up triangles) and the VdW interaction (down triangles), for $\U/\Ec=2$, $10$, and $40$ [the markers have the same color scales as in (a)].
}
\label{fig:strong_coupling}
\end{figure}

The breakdown of the Bogoliubov approach signals the onset of the strongly-correlated regime.
For 1D systems, this universally happens in the low-density sector.
This is visible on Fig.~\ref{fig:structurefactor}(b) for roughly $n\lesssim 0.5\Rc^{-1}$ at $\U=5\Ec$ and $\U=10\Ec$, and for all considered densities for $\U=20\Ec$ and $\U=40\Ec$.
More precisely, it occurs below the line at $K=1$ in Fig.~\ref{fig:phase_diagram}.
For vanishingly small densities, the vacuum-superfluid transition of bosons falls in the same universality class of free fermions~\cite{sachdev2001}, and shows a characteristic $n \propto \sqrt{\mu}$ behaviour, see Fig.~\ref{fig:eqnofstates}.
For larger densities and sufficiently close to the line at $K=1$, the gas behaves as HRs~\cite{sutherland1971groundstate,sutherland1971quantum,wadati2002one,vsamaj2008introduction} or super-Tonks-Girardeau gas~\cite{astrakharchik2005beyond}, characterized by $K\approx (1-na_\mathrm{1D})^2$, where $a_\mathrm{1D}$ is the scattering length.
In the $n\to 0$ limit, we recover the Tonks-Girardeau gas, while, for increasing $n$, the value of $K$ decreases or increases, depending on the sign of $a_\mathrm{1D}$.
For Rydberg-dressed interaction, the scattering length turns from negative to positive values at $\U \approx 1.09\Ec$~\cite{teruzzi2017microscopic}.
For larger $\U$, the values of $K$ show clear HR-like behavior in the low enough density limit, see Fig.~\ref{fig:strong_coupling}(a).
Moreover, this behavior is preserved at increasingly large values of $n a_{\mathrm{1D}}$, the stronger the Rydberg-dressed interaction core $\U$.
Finally, strong enough interactions and low densities, the interaction core is not probed by scattering particles and the Rydberg-dressed gas behaves as with van der Waals (VDW) interactions with the same constant $C_6=\U\Rc^6$, see Fig.~\ref{fig:strong_coupling}(b).

The consequences of HR-like behavior are also evident in Fig.~\ref{fig:structurefactor}(b).
Focusing on the lowest densities for each value of $\U$, which always fall within the strongly-correlated regime, we observe a sharp peak at $q = 2\kF$ across all values of $\U$. This peak becomes increasingly prominent as $\U$ increases, a characteristic feature of HR systems with $K < 1$.
This behavior can be understood from the Fourier transform relationship between the pair correlation function $g_2(x)$ and the static structure factor $S(q)$~\cite{mazzanti2008ground,motta2016dynamical,yu2026thermodynamics}.
In particular, the power-law decay of $g_2(x)$ in real space translates into a singular structure in momentum space at $2\kF$, which increases with the interaction strength $\U$. This trend can be attributed to the growth of the interaction strength, which enhances correlations and amplifies density modulations at the corresponding wave number.
When $K < 1/2$, the peak at $q = 2\kF$ exhibits genuine divergence. Physically, such a divergent peak signals the emergence of strong solid-like ordering, albeit with quasi-long-range rather than true long-range order, characterized by algebraically decaying two-body correlations. Since the system remains 1D superfluid, the observed behavior is appropriately described as a quasi-supersolid.

\subsection{Coexistence of roton-driven ordering and particle-hole backscattering-induced ordering}
Consider now the intermediate regime, where both roton-driven density correlations and particle-hole backscattering ordering processes coexist.
First, let us fix the interaction strength and increase the density, which essentially corresponds to a crossover from the strongly-correlated regime to the meanfield regime.
Take, for instance, the third column in Fig.~\ref{fig:structurefactor}, which corresponds to $\U=20\Ec$.
For low density, the QMC result shows a strong peak at $q=2\kF$ induced by particle-hole backscattering processes but no sign of the roton predicted by meanfield theory.
When the density decreases, the sharp peak at $q=2\kF$ progressively decreases and a smooth local maximum associated to the roton appears and grows, although at a momentum significantly shifted with respect to the meanfield prediction.

Second, let us fix the density.
As the interaction strength increases, the roton mode is enhanced in the Bogoliubov regime and the quasisolidity peak at $q=2\kF$ is enhanced in the strongly-correlated regime.
However, it is interesting to see the intermediate regime.
For instance, for densities around $0.8\Rc^{-1}$, we find that strong-enough interactions overwhelm the roton-driven density correlation, although the latter is enhanced as well.
Especially, the roton gap should be closed at this density for $\U=40\Ec$ in meanfield theory, but this is prevented by the strong quasisolidity.
Hence, the TLL behavior is kept even in the symmetry breaking phase in the meanfield theory.

\section{Beyond Tomonaga-Luttinger liquids and clustering} \label{sec:beyondTLL}
When the roton momentum becomes comparable to the Fermi momentum $\kF$, their tendencies toward crystallization reinforce each other.
As discussed in Ref.~\cite{rossotti2017quantum}, cluster nucleation may arise when the minimum of $\tilde{V}(q)$ is commensurate with $\kF$ in the strongly interacting regime.
On the other hand, for sufficiently strong interactions and high densities, our QMC results reveal deviations from TLL behavior, as evidenced by discrepancies among the Luttinger parameter estimates, see Fig.~\ref{fig:K_normalTLL}(d) for $\U = 40\Ec$ and $n\Rc \gtrsim 0.8$.
In this section, we show that beyond-TLL behavior is a precursor to clusterization, with the cluster phase emerging also out of incommensurability across a broader parameter range.

\subsection{Beyond Tomonaga-Luttinger liquids}
\label{sec:phenomena_beyondTLL}
\begin{figure}[t]
\centering
\includegraphics[width=\linewidth]{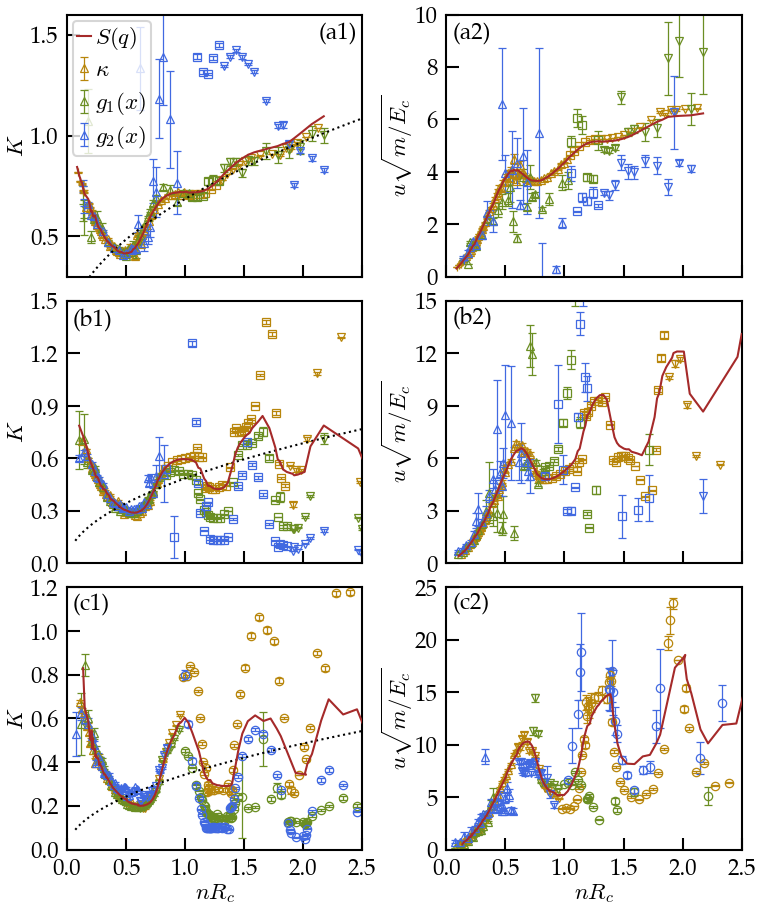}
\caption{(a1)-(c1) Luttinger parameter $K$; (a2)-(c2) speed of sound $u$.
The values of $\U/\Ec$ are $10$, $20$, and $40$ from~(a) to (c).
The figures from (a1) to (c2) share the same set of legends.
The errorbars of $S(q)$ are sufficiently small such that we can discard them and connect the points directly.
In (a1)-(c1), the dotted lines indicate the Bogoliubov prediction.
}
\label{fig:fits}
\end{figure}
\begin{figure*}[t]
\centering
\includegraphics[width=\linewidth]{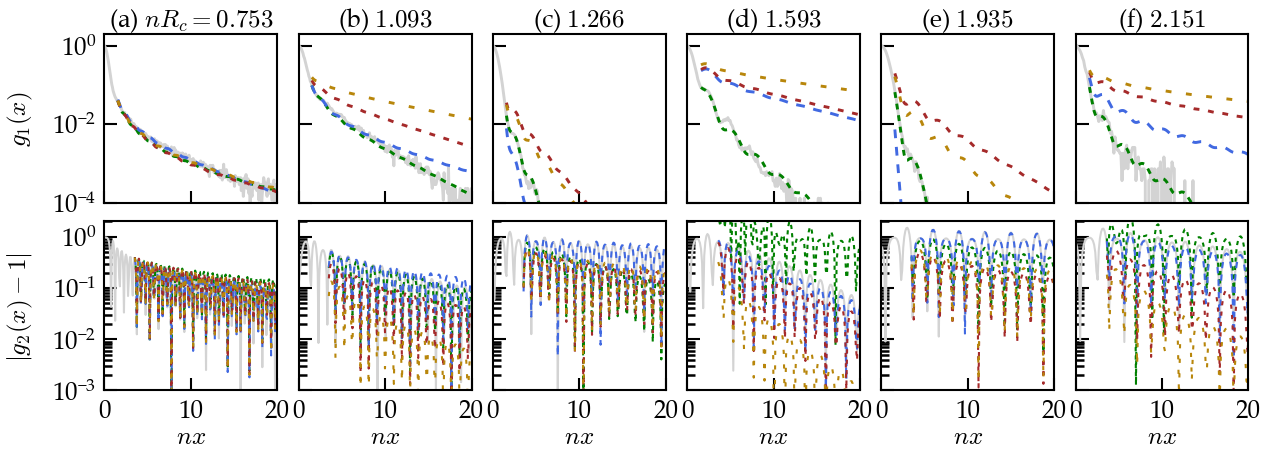}
\caption{
Correlation functions and fits at $\U=40\Ec$ for various densities as titled.
They correspond to normal TLL, beyond-TLL, 2-particle cluster, beyond-TLL, 3-particle cluster, beyond-TLL regimes, from left to right.
They also correspond to the red triangles in Fig.~\ref{fig:phase_diagram}.
Solid gray lines are QMC results.
Green, blue, red, and orange dashed lines are obtained from the fit to $g_1$ or $g_2$ by keeping all fitted parameters unchanged except for $K$, which is replaced by the value estimated from $g_1$, $g_2$, $S(q)$, and $\kappa$ respectively.}
\label{fig:fits_U40}
\end{figure*}
Results for the various estimates of the Luttinger parameter $K$ and, correspondingly of the speed of sound $u$, are shown in Figs.~\ref{fig:fits}(a)--(c), for $\U=10$, $20$, and $40E_c$ from top to bottom, covering a broader range of parameters compared to Fig.~\ref{fig:K_normalTLL}.

For all values of $\U$, we find good agreement among the four estimates of $K$ for $n\lesssim 0.8\Rc^{-1}$, with $K<1$, indicating that the system is in the strongly-correlated TLL regime.
At higher densities, discrepancies appear among the four estimates of $K$.
As discussed above, fits to $g_2(x)$ can be delicate and may fail in the TLL regime. For this reason, we disregard estimates of $K_{g_2}$ for identifying breakdown of TLL behavior.
Note, however, that we generally obtain good fits of $g_2(x)$ in the cluster phases identified below.
In contrast, the fits of $g_1(x)$ and $S(q)$ are generally good, while $K_\kappa$ is obtained independently of fits.
Therefore, when these reliable estimates of $K$ become mutually inconsistent, we interpret this as evidence for beyond-TLL behavior.
Based on this analysis, we identify beyond-TLL behavior in the regions $0.91 \lesssim n\Rc \lesssim 1.46$ and $n\Rc \gtrsim 1.52$ for $\U=20\Ec$, and in the region $n\Rc \gtrsim 0.97$ for $\U=40\Ec$.
Proceeding similarly for various values of $\U$, we identify the beyond-TLL regime delimited by the dashed blues line in Fig.~\ref{fig:phase_diagram}.

To further assess that the various quantities studied above are indeed governed by incompatible Luttinger parameters, we plot in Fig.~\ref{fig:fits_U40} the Haldane formulas for $g_1(x)$ [Eq.~(\ref{eq:TLL_g1}), upper row] and $g_2(x)$ [Eq.~(\ref{eq:TLL_g2}), lower row] using the four estimates of $K$.
The results shown in Fig.~\ref{fig:fits_U40} correspond to points marked by the orange, blue, and green triangles near the right-hand-side edge of Fig.~\ref{fig:phase_diagram}.
In Fig.~\ref{fig:fits_U40}, the solid gray lines denote the QMC results, while the dashed lines denote the Haldane formulas using
$K_{S(q)}$~(red).
$K_{\kappa}$~(orange).
$K_{g_1}$~(green).
and $K_{g_2}$~(blue).
For $g_1(x)$, the agreement between the QMC data and the fitted Haldane formula~(\ref{eq:TLL_g1}), using $K_{g_1}$, which is directly extracted from this fit, is generally good.
For $g_2(x)$, the agreement between the QMC data and the fitted Haldane form~(\ref{eq:TLL_g2}), using $K_{g_2}$, which is directly extracted from this fit,  is also mostly satisfactory.
Nevertheless, visible deviations appear in (b), (d), and (f), all of which lie in the beyond-TLL regimes.
More importantly, when the other estimates of $K$ are inserted into the Haldane formulas, clear differences among them are observed, except for panel~(a), which lies in the normal TLL regime.
The deviations observed for all other densities considered in Fig.~\ref{fig:fits_U40} confirm the breakdown of TLL behavior.
Note that, in cluster phases, the short valid distance of $g_1(x)$ cannot assert the agreement with Haldane's formula in the asymptotic limit.
Nevertheless, the mismatch among all estimates is clear.
Hence, it is definitely beyond the description of TLL theory.

\subsection{Cluster phases}
\label{sec:cluster}

We now turn to the cluster phases, which lie inside the beyond-TLL regime.

\subsubsection{Identification of cluster phases}
\begin{figure}[b]
\centering
\includegraphics[width=\linewidth]{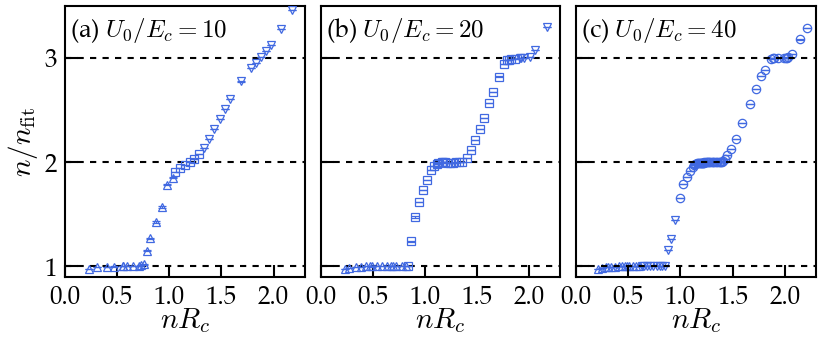}
\caption{Estimates of $n/n_\mathrm{fit}$ for various $\U$'s, obtained from fits of $g_2(x)$.
Error bars are smaller than the markers.}
\label{fig:nfit}
\end{figure}

It is insightful to examine the behavior of the fitted density via the pair correlation function $g_2(x)$. While the Haldane formula breaks down in the beyond-TLL regime, the fitted density $n_\mathrm{fit}$ accurately reproduces the oscillation frequency, even within this regime, see all lower panels in Fig.~\ref{fig:fits_U40}. Furthermore, the fits of $g_2(x)$ are actually robust in the cluster phase, which we study here. Estimates of the ratio $n/n_\mathrm{fit}$ versus density are shown in Fig.~\ref{fig:nfit} for various values of $\U$. At low density, $n/n_\mathrm{fit}$ remains close to unity across all values of $\U$, except at the lowest densities where finite-temperature effects become apparent.
This low-density behavior aligns with the standard TLL behavior.
As the density increases, $n/n_\mathrm{fit}$ deviates from unity for $n\Rc \gtrsim 0.8$ and all three $\U$ values considered.
This indicates that $g_2(x)$ oscillates spatially at a frequency exceeding the expectation one based on the gas density for a TLL, a hallmark of beyond-TLL behavior.
Notably, the precise nature of this deviation depends strongly on the value of $\U$.

For $\U=10\Ec$, the ratio $n/n_\mathrm{fit}$ increases monotonically. However, for $\U=20\Ec$ and $\U=40\Ec$, it exhibits distinct plateaus at integer values, $n/n_\mathrm{fit}=2$ around $n\Rc \simeq 1.4$ and $n/n_\mathrm{fit}=3$ around $n\Rc \simeq 2$. This suggests that the dominant harmonic of $g_2(x)$ is $q=\kF$, rather than the $q=2\kF$ expected for a TLL. Such behavior aligns with the observed peaks in $S(q)$ at $n\Rc = 1.426$ for both $\U=20\Ec$ and $\U=40\Ec$, see Figs.~\ref{fig:structurefactor}(b3) and (b4).
The divergent peaks in $S(q)$, together with $K_{g_2}\leqslant 0.1$ [see Figs.~\ref{fig:fits}(b1) and (c1)], further suggest that these cluster states have a strong tendency towards quasi-supersolid order.
These findings are also consistent with the two-particle cluster phase reported in Ref.~\cite{rossotti2017quantum}.
Our results show that the effective spatial period of the system becomes an integer multiple of $n^{-1}$, indicating that particles are grouped into molecule-like clusters.
Additionally, we find that the cluster phase is stabilized over extended ranges of chemical potential or density, as evidenced by the plateaus.

We identify a stable cluster phase when an integer plateau of $n/n_\mathrm{fit}$ is present.
To locate the boundaries of each cluster phase, we find the density where $n/n_\mathrm{fit} = \sigma \pm 0.01$.
Repeating this analysis for various values of $\U$ and interpolating, we plot the cluster phases at $\sigma=2$ and $\sigma=3$ (solid green lines) in the phase diagram of Fig.~\ref{fig:phase_diagram}.
The red star denotes the critical point reported in Ref.~\cite{rossotti2017quantum}, which consistently lies in our commensurate cluster phase at $\sigma=2$. In fact, we find that it is located on the boundary of a larger cluster lobe.

\subsubsection{Cluster-TLL behavior}
In the cluster phases, the rearrangement of particles invalidates the ordinary TLL description based on individual particles.
The large sound velocities obtained in Figs.~\ref{fig:fits}(b2) and (c2), as well as the diverging peaks of $S(q)$ visible in Figs.~\ref{fig:structurefactor}(b3) and (b4), are also characteristic of a stiff, solid-like behaviour.
The third and fifth columns of Fig.~\ref{fig:fits_U40}, which correspond to the $\sigma=2$ and $\sigma=3$ particle cluster phases, respectively, clearly show beyond normal TLL behavior.

It has been shown that the pair distribution function $g_2(x)$ in the cluster phases can instead be described by a cluster TLL theory~\cite{mattioli2013cluster,dalmonte2015cluster,rossotti2017quantum}.
Since the clusters are stable and difficult to break, they should be treated as new effective particles with mass $M_\mathrm{clu}=\sigma M$.
and the low-energy excitations are expected to be governed by the collective motion of these clusters.

\begin{figure}[t!]
\centering
\includegraphics[width=\linewidth]{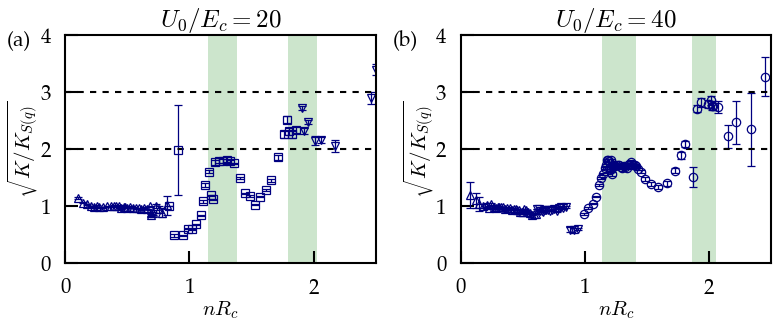}
\caption{Cluster-TLL ratio $\sqrt{K_{S(q)}/K_{g_2}}$ versus density for $\U=20\Ec$ (let) and $\U=40\Ec$ (right). Green shaded regions indicate the two- and three-particle cluster phases identified from the plateaus of $n/n_\mathrm{fit}$.}
\label{fig:Kratio}
\end{figure}

The particle density operator can then be expressed in terms of the fields of clusters, basically $\sigma$ times the cluster density operator, and
the large-distance behavior of the particle pair correlation function may be written
\begin{equation} \label{eq:cluster_g2}
g_2(x) = 1 - \frac{2\sigma^2 K_\mathrm{clu}}{\left[2\pi nd(x)\right]^2} + \sum_{m =1}^\infty A_m \frac{\cos
\left( \frac{2\pi mn}{\sigma} x \right)}{\left[nd(x)\right]^{2m^2K_\mathrm{clu}}}.
\end{equation}
Since $K_\mathrm{clu}$ is small in the cluster phases, 
the first harmonic term ($m=1$) dominates, 
so that fitting $g_2(x)$ with the ordinary Haldane form effectively gives $K_{g_2}\approx K_\mathrm{clu}$.
In contrast, the static structure factor at small wave number is dominated by the non-oscillating term and we expect
$S(q) \simeq \frac{\sigma^2K_\mathrm{clu}q}{2\kF}$.
Hence, we expect
$K_{S(q)} \simeq \sigma^2 K_\mathrm{clu}$,
which yields
the modified relation between $K_{S(q)}$ and $K_{g_2}$
\begin{equation}
\sqrt{\frac{K_{S(q)}}{K_{g_2}}} \approx \sigma.
\end{equation}
This cluster-TLL prediction is checked in Fig.~\ref{fig:Kratio}, where we show $\sqrt{K_{S(q)}/K_{g_2}}$ versus density for $\U=20\Ec$ and $\U=40\Ec$.
In the commensurate cluster phases, identified by the plateaus of $n/n_\mathrm{fit}$ and indicated by the green shaded regions, the values are close to the corresponding integer $\sigma$, in agreement with the cluster TLL prediction.
This confirms that the low-energy physics of the cluster states is described by a cluster TLL rather than by an ordinary TLL of individual particles.

\subsubsection{Statistics of clusters}

\begin{figure}[t]
\centering
\includegraphics[width=\linewidth]{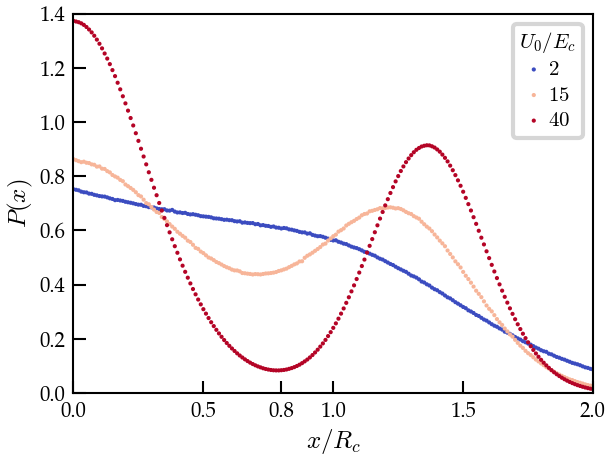}
\caption{Distribution of inter-particle distances at $n=1.25\Rc^{-1}$, with various values of $\U$.
They correspond to the normal, boundary, and deep two-particle cluster phases, respectively.
Error bars are smaller than the markers.}
\label{fig:spacing_distribution_n1.25}
\end{figure}

We then study the spatial distribution of particles.
The difference between the cluster phases and the normal TLL phase can be revealed by examining the distribution of nearest-neighbor distances.
We do this by leveraging the path-integral representation used in QMC, the world lines of which directly yield the particle positions.
This distribution is shown in Fig.~\ref{fig:spacing_distribution_n1.25} for density $n=1.25\Rc^{-1}$ and several values of $\U$,
corresponding to the normal TLL regime ($\U=2\Ec$),
the two-particle cluster phase ($\U=40\Ec$),
and close to the boundary between them for ($\U=15\Ec$).

In the normal TLL regime ($\U=2\Ec$, blue line), the distribution is smooth and decreases monotonically, with a typical width of about $1.5\Rc$.
In contrast, in the cluster phase ($\U=40\Ec$, red line), the distribution shows two well-separated Gaussian-like peaks,
centered at $x_1=0$ and $x_2\simeq 1.4\Rc$, respectively, with a width roughly equal to $0.3\Rc$ each.
The minimum between the peaks is located around $x_m\simeq 0.8\Rc$.
The integrals over the intervals $[0,0.8\Rc)$ and  $[0.8\Rc, 2.0\Rc)$ are approximately one half each.
This bimodal distribution, also visible at the cluster-phase boundary ($\U=15\Ec$, orange line),
strongly favours the clustering picture, where
the first and second peaks correspond to nearest-neighbor separations within clusters and between neighboring clusters, respectively.

\begin{figure}[t]
\centering
\includegraphics[width=\linewidth]{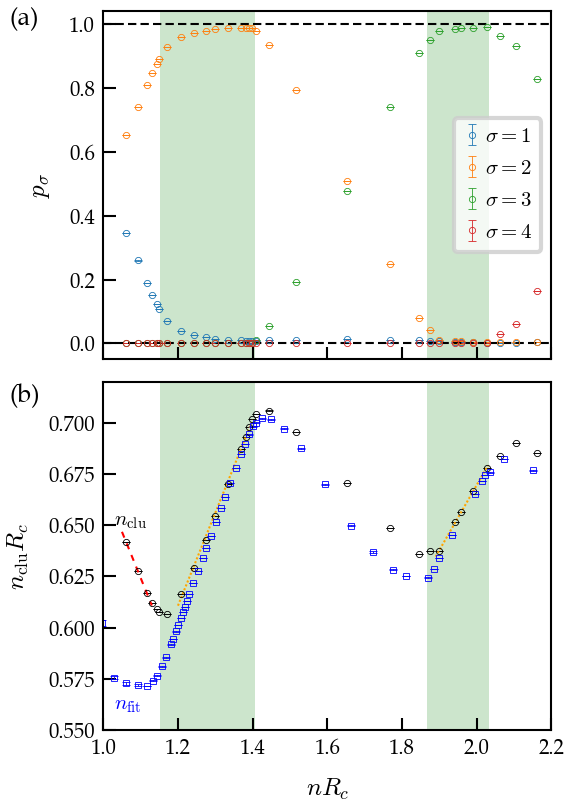}
\caption{
Cluster statistics around the 2- and 3-particle cluster phase for $\U=40\Ec$ and $T=0.4\Ec/\kB$.
Green shaded regions indicate the commensurate 2- and 3-particle cluster phases, respectively.
(a)~Proportion of each type of cluster.
(b)~Cluster density as a function of particle density.
In panel~(b), black circles denote $n_\mathrm{clu}$, while blue squares denote $n_\mathrm{fit}$ obtained from the fits of $g_2(x)$.
In most cases, the shown error bars are smaller than the symbols.
The dashed lines are linear fits:
the red line is $n_\mathrm{clu}=-0.45n+1.12$,
and the orange lines are $n_\mathrm{clu}=0.454n+0.066$ and $n_\mathrm{clu}=0.318n+0.034$, for $\sigma=2$ and $\sigma=3$, respectively.
}
\label{fig:cluster_density}
\end{figure}

We further study the statistics of clusters directly from the QMC configurations.
Clustering is induced by the fact that the soft-core potential is nearly flat at short distances.
As a result, particles do not strongly repel each other when their separation is very small.
We therefore introduce a threshold cluster diameter of $0.8\Rc$, which approximately corresponds to the minimum in Fig.~\ref{fig:spacing_distribution_n1.25}.
Using this threshold, we group particles into clusters and count both the total number of clusters and the number of particles contained in each cluster.
More specifically, we first choose a particle as a starting point.
Assuming the position of the particle is $x_0$,
all particles within the interval $[x_0,x_0+0.8\Rc)$, including the starting particle, are identified as belonging to the same cluster.
We then move to the next unassigned particle and repeat the procedure until every particle has been assigned to a cluster.
The number of clusters containing $\sigma$ particles is denoted by $N_\sigma$, and the total number of clusters is
\begin{equation}
N_\mathrm{clu}=\sum_\sigma N_\sigma .
\end{equation}
This geometrical identification method should, however, be interpreted with care.
It is meaningful mainly near the clustered regimes, where particles are expected to form well-defined local groups.
In the meanfield or noncluster regimes, the same procedure may still produce a nonzero count of $\sigma$-particle clusters, simply because particles can occasionally be close to one another.
Such counts should not be interpreted as evidence for physical cluster formation.

Figure~\ref{fig:cluster_density} shows the proportion of each type of cluster, 
\begin{equation}
p_\sigma=\frac{N_\sigma}{N_\mathrm{clu}},
\end{equation}
together with the cluster density, $n_\mathrm{clu}=N_\mathrm{clu}/L$, across the two- and three-particle cluster regimes.
Figure~\ref{fig:cluster_density}(a) directly visualizes the process of cluster formation as the density increases.
Across the transition from the normal phase to the two-particle cluster phase, $p_{\sigma=1}$ decreases toward zero and $p_{\sigma=2}$ increases toward unity, while the proportion of other clusters are vanishingly small.
The value of $p_{\sigma=2}$ remains close to unity throughout the two-particle cluster phase.
The remaining small nonzero value of $p_1$ at the entrance of the two-particle cluster phase is probably due to thermal excitations or the disassociation of clusters.
When the density further increases out of the two-particle cluster phase, $p_{\sigma=2}$ decreases while $p_{\sigma=3}$ increases.
In the three-particle cluster phase, $p_{\sigma=3}$ becomes close to unity, and the other components are nearly suppressed.
This confirms the strong clusterization of all particles in the corresponding cluster phases.

The total cluster density, shown in Fig.~\ref{fig:cluster_density}(b), provides a complementary characterization.
In the commensurate cluster phases, we find that $n_\mathrm{cluster}$ is approximately equal to $n/2$ and $n/3$ in the two- and three-particle cluster phases, respectively.
This is consistent with the fact that adding two or three particles to the system creates approximately one additional cluster in the corresponding commensurate phase.
Away from the commensurate phases, however, the cluster density decreases on both sides.
The physical interpretation of this behavior is less straightforward.
This decrease suggests that adding particles can induce a rearrangement in which particles tend to approach each other so that some existing cluster can merge or reorganize, thereby reducing the total cluster density.

\section{Conclusions}\label{sec:conclusions}
In this work, we have investigated 1D bosons interacting via a Rydberg-dressed soft-core potential using continuous-space path-integral quantum Monte Carlo simulations.
The finite range of the interaction and the negative part of its Fourier transform generate a roton mode roughly around the finite momentum $q^*$, 
while particle-hole backscattering processes in 1D naturally favour density waves at twice the Fermi momentum $2\kF$.
The competition of both corresponding length scales gives rise to a rich phase diagram.
By varying the interaction strength and density, 
we have mapped out this phase diagram and identified three main regimes: 
an ordinary TLL regime, 
a regime where the ordinary single-particle TLL description breaks down, 
and commensurate cluster phases.

In the ordinary TLL regime, the low-energy physics is consistently described by a single Luttinger parameter $K$, as confirmed by the agreement between independent estimates extracted from the compressibility, the low-momentum structure factor, and correlation functions.
Within this regime, the system crosses over from a Lieb-Liniger-like region with $K>1$, to a HR-like region with $1/2<K<1$, and finally to a strongly-correlated regime with $K<1/2$.
In the latter case, the static structure factor develops algebraically diverging peaks at $q=2\kF$, which is characteristic of solid-like quasi-long-range diagonal order,
while one-body correlations still decay algebraically, which is characteristic of quasi-long range off-diagonal superfluid order.
In the context of 1D systems, this regime may be viewed as a quasi-supersolid regime.
This regime appears when the roton and the particle-hole backscattering processes have comparable characteristic momenta, \ie~$q^* \sim 2\kF$.
Besides, comparison of the QMC static structure factor with the Bogoliubov-Feynman prediction also shows that the validity of the meanfield description is not controlled solely by the value of $K$,
but also by finite-momentum correlations associated with the roton minimum and by the proximity to clustering.

For stronger interactions and higher densities, 
we find clear deviations from the ordinary particle TLL description.
These deviations are signaled by incompatible values of the Luttinger parameter extracted from different observables, 
and by a change in the oscillation period of the pair correlation function.
In particular, the ratio between the particle density and the fitted oscillation density develops integer plateaus, indicating that particles bind into effective clusters containing a fixed number $\sigma$ of particles.
The two- and three-particle cluster phases are further supported by the linear parts
in the equation of state, the strong peaks in the static structure factor, and the statistics of inter-particle distances.
In these phases, the appropriate low-energy degrees of freedom are not individual particles but clusters, 
whose collective motion is consistent with an emergent cluster-TLL description.
Clustering implies that the effective Fermi momentum is divided by the number of particles in each cluster $\sigma$  and the momentum of the particle-hole backscattering processes becomes $2\kF/\sigma$.
Its interplay with the roton around $q^*$ induces strong peaks in the structure factor around these values, qualitatively similar to the process leading to quasi-supersolid behaviour in the absence of clustering.

Our results show that the roton-gap closing scenario in the meanfield theory expected in higher dimensions is strongly reshaped by 1D correlations. 
Rather than producing true long-range supersolid order, 
the interplay of roton-induced and particle-hole backscattering-induced tendencies to formation of density waves in 1D systems gives rise to quasi-long-range density correlations, ordinary TLL behavior, and commensurate cluster phases.
The cluster phases appear when the roton length scale becomes comparable with the density of effective clusters, providing a microscopic mechanism for the breakdown of the ordinary single-particle TLL.

Several questions remain open.
In particular, the nature of the beyond-TLL regions between ordinary TLL behavior and commensurate cluster phases deserves further study.
It would also be valuable to characterize the excitation spectrum directly, where separate particle-like and cluster-like branches may emerge. 
More generally, extensions of Bogoliubov theory, cluster effective theories, or impurity-inspired approaches may provide complementary analytical descriptions of the crossover from roton-enhanced correlations to cluster formation. 
The phenomena discussed here should be relevant not only for Rydberg-dressed gases, but also for other 1D systems with soft-core or van der Waals interactions in elongated traps.

\begin{acknowledgments}
We thank Zhaoxuan Zhu for fruitful discussions.
We acknowledge the CPHT computer team for valuable support.
This research was supported by
the Agence Nationale de la Recherche under projects QuanTEdu-France (No. ANR-CMAQ-002 France 2030) and QUTISYM (No. ANR-23-PETQ-0002),
the IPParis Doctoral School, 
and HPC/AI resources from GENCI-TGCC (Grants AD010510300R1 and A0200510300) using the ALPS scheduler library and statistical analysis tools~\cite{troyer1998,ALPS2011}.
\end{acknowledgments}

\section*{Data availability}
The data that support the findings of this article are not
publicly available. The data are available from the authors
upon reasonable request

\bibliographystyle{revtexlsp}
\bibliography{bibYSJ, biblioLSP}

\end{document}